\documentclass[aps,prx,longbibliography,superscriptaddress,twocolumn,showpacs]{revtex4-2}
\usepackage[utf8]{inputenc}
\usepackage[american,british]{babel}
\usepackage[T1]{fontenc}

\usepackage{float}
\usepackage{amssymb}

\usepackage{graphicx}
\usepackage{xcolor}
\usepackage{dcolumn}
\usepackage{bm}
\usepackage[normalem]{ulem}

\usepackage{amsmath,amsthm,amssymb,mathrsfs,mathtools,amsfonts,braket}
\usepackage[makeroom]{cancel}
\usepackage[caption=false]{subfig}

\usepackage[colorlinks=true, linktocpage=true]{hyperref}
\hypersetup{
    allcolors = {black}
}

\usepackage{braket}
\usepackage{todonotes}

\usepackage[Option]{overpic} 
\usepackage{mathdots}
\usepackage{dsfont}
\usepackage{mathrsfs}
\DeclareMathAlphabet{\mathscrbf}{OMS}{mdugm}{b}{n}

 \newcommand{\RN}[1]{
  \textup{\expandafter{(\romannumeral#1)}}%
}
\usepackage{comment}

\usepackage{bigints}
\usepackage{float}
\usepackage{scalerel,stackengine,amsmath}

\begin{document}
\title{Equilibrium Quantum Impurity Problems \\ via Matrix Product State Encoding of the Retarded Action}

\author{Benedikt Kloss}
\thanks{These authors contributed equally to this work.\\
Correspondence should be addressed to:\\
Benedikt Kloss (bene.kloss@gmail.com) and\\
Julian Thoenniss (julian.thoenniss@unige.ch).}
\affiliation{Center for Computational Quantum Physics, Flatiron Institute, New York, New York 10010, USA}

\author{Julian Thoenniss}
\thanks{These authors contributed equally to this work.\\
Correspondence should be addressed to:\\
Benedikt Kloss (bene.kloss@gmail.com) and\\
Julian Thoenniss (julian.thoenniss@unige.ch).}
\affiliation{Department of Theoretical Physics,
University of Geneva, 30 Quai Ernest-Ansermet,
1205 Geneva, Switzerland}

\author{Michael  Sonner}
\affiliation{Department of Theoretical Physics,
University of Geneva, 30 Quai Ernest-Ansermet,
1205 Geneva, Switzerland}

\author{Alessio Lerose}
\affiliation{Department of Theoretical Physics,
University of Geneva, 30 Quai Ernest-Ansermet,
1205 Geneva, Switzerland}

\author{Matthew T. Fishman}
\affiliation{Center for Computational Quantum Physics, Flatiron Institute, New York, New York 10010, USA}

\author{E. M. Stoudenmire}
\affiliation{Center for Computational Quantum Physics, Flatiron Institute, New York, New York 10010, USA}

\author{Olivier Parcollet}
\affiliation{Center for Computational Quantum Physics, Flatiron Institute, New York, New York 10010, USA}
\affiliation{Universit\'e Paris-Saclay, CNRS, CEA, Institut de Physique Th\'eorique, 91191, Gif-sur-Yvette, France}

\author{Antoine Georges}
\affiliation{Center for Computational Quantum Physics, Flatiron Institute, New York, New York 10010, USA}

\affiliation{Coll\`ege de France, 11 Place Marcelin Berthelot, 75005 Paris, France}
\affiliation{CPHT, CNRS, \'Ecole Polytechnique, Institut Polytechnique de Paris, Route de Saclay, 91128 Palaiseau, France}
\affiliation{Department of Quantum Matter Physics, University of Geneva,
24 Quai Ernest-Ansermet, 1205 Geneva, Switzerland}

\author{Dmitry A. Abanin}
\affiliation{Department of Theoretical Physics,
University of Geneva, 30 Quai Ernest-Ansermet,
1205 Geneva, Switzerland}
\affiliation{Google Research, Mountain View, CA, USA}

\date{\today}

\begin{abstract}
In the $0+1$ dimensional imaginary-time path integral formulation of quantum impurity problems, the retarded action encodes the hybridization of the impurity with the bath. In this Article, we explore the computational power of representing the retarded action as matrix product state (RAMPS).
We focus on the challenging Kondo regime of the single-impurity Anderson model, where non-perturbative strong-correlation effects arise at very low energy scales. We demonstrate that the RAMPS approach reliably reaches the Kondo regime for a range of interaction strengths $U$, with a numerical error  scaling as a weak power law with inverse temperature. We investigate the convergence behavior of the method with respect to bond dimension and time discretization by analyzing the error of local observables in the full interacting problem and find polynomial scaling in both parameters. 
Our results show that the RAMPS approach offers promise as an alternative tool for studying quantum impurity problems in regimes that challenge established methods, such as multi-orbital systems. Overall, our study contributes to the development of efficient and accurate non-wavefunction-based tensor-network methods
for quantum impurity problems.
\end{abstract}

\maketitle

\section{Introduction}
 An accurate theoretical description of a general quantum impurity coupled to a bath of itinerant fermions across a wide range of energy scales remains a major challenge to date, even in thermal equilibrium\,\cite{Anderson61Localized,hewson_1993}. At the same time, such quantum impurity models (QIMs) play a central role in modern condensed matter physics: They are interesting {\it per se} for the study of emergent strong-correlation phenomena, such as the Kondo effect\,\cite{Wilson75Renormalization,hewson_1993}, and form the foundation of powerful quantum embedding techniques, such as dynamical-mean-field theory (DMFT)\,\cite{Georges92Hubbard,Georges96Dynamical,Kotliar06Eletronic}. This makes them a crucial topic of study for a broad range of applications. \\

As the exponential number of parameters of the full many-body problem prevents an exact treatment, sophisticated numerical techniques are required to solve QIMs. In past decades, a wide range of different algorithms for QIMs has arisen, including approaches based on exact diagonalization\,\cite{Caffarel94Exact,Koch08Sum,Mejuto-Zaera20Efficient}, matrix product states (MPS)\,\cite{Wolf14Chebyshev,Garcia04Dynamical,Nishimoto06Dynamical,Bauernfeind17Fork,Weichselbaum09Variational,Werner23Configuration}, configuration interaction expansions\,\cite{Zgid12Truncated,Werner23Configuration}, Markov-chain Monte Carlo\,\cite{Hirsch86Monte}, continuous-time Quantum Monte Carlo (CT-QMC)\,\cite{Rubtsov2005continuous,Gull2008continuous,Werner06Hybridization,Werner06continuous,Muhlbacher08Real,Gull11Numerically,Cohen15Taming,Eidelstein20Multiorbital,Parcollet15Triqs,Seth26Triqs/CTHYB,Shinaoka17Continuous,Shinaoka20Efficient}, and the numerical renormalization group (NRG)
\,\cite{Wilson75Renormalization,Bulla08Numerical,Lee16Adaptive,Lee17Doublon}.  The latter two classes of algorithms have gained special importance for state-of-the-art implementations of QIM solvers in thermal equilibrium\,\cite{Lee17Doublon,Seth26Triqs/CTHYB}.\\

In NRG, a sequence of effective Hamiltonians, each of which describes the low-energy physics of the system at successively smaller energy scales, is iteratively diagonalized. While a logarithmic energy discretization leads to an excellent accuracy at low energy scales which allows to resolve signatures of Kondo physics with high precision, NRG may be less accurate in higher energy parts of the spectrum.\\

In contrast, CT-QMC does not rely on an explicit Hamiltonian representation of the bath. Instead, it is based on Monte Carlo sampling of the imaginary-time Green's function or the partition sum using a perturbative expansion in the hybridization function between impurity and bath. While CT-QMC is in principle numerically exact, it typically suffers from the fermionic sign problem for impurity--bath couplings that are not diagonal in spin space.\\

Tensor network state methods have traditionally relied on a Hamiltonian formulation of the problem, requiring a bath discretization and time evolution of the complete system's wavefunction. Together with limited accessible timescales on the real time axis due to entanglement growth, their accuracy at low frequencies is limited. Tensor network state methods can also be applied to compute Green's functions on the imaginary-time axis, where the entanglement growth is lessened but still challenging for realistic models\,\cite{Wolf2015b,PhysRevB.101.041101,PhysRevB.105.195107}.
Against this background, the development of new approaches to accurately and efficiently solving QIMs remains an important endeavor.\\

Recently, several non-wavefunction-based tensor network approaches have emerged. This includes the exact evaluation of a perturbative expansion to very high orders aided by tensor network state compression\,\cite{erpenbeck2023tensor,Fernandez22Learning} as well as expression of the hybridization between impurity and bath via the Feynman-Vernon influence functional encoded as a matrix product state (MPS) in the temporal domain\,\cite{thoenniss2022nonequilibrium,thoenniss2022efficient,ng2023real}. We note that techniques closely related to the latter have been developed first for bosonic baths or interacting spin chains\,\cite{PhysRevLett.102.240603,PhysRevX.11.021040,SONNER2021168677,strathearn2018efficient,10.1063/5.0047260} and have only recently been generalized to fermionic particles.\\

In this Article, we explore such a non-perturbative tensor-network approach to equilibrium QIMs defined by the retarded action (RA) and the local impurity Hamiltonian. By constructing an efficient encoding of the RA as MPS (RAMPS), we can accurately compute arbitrary local impurity observables by means of efficient tensor network contractions. While some of the authors have previously applied a similar approach to nonequilibrium QIMs\,\cite{thoenniss2022efficient,thoenniss2022nonequilibrium}, where it has demonstrated competitiveness in accuracy and efficiency compared to state-of-the-art methods, here we extend it to equilibrium QIMs. \\

Here, we focus on technical aspects and proof-of-principle calculations. We use the single-impurity Anderson model (SIAM) to assess the accuracy of the algorithm in the Kondo regime as a function of numerical and physical parameters. To this end, we provide benchmarks against analytical and numerically exact (CT-QMC) results in the noninteracting and interacting case, respectively. \\

Our findings indicate that this approach is a promising technique for accurately and efficiently solving QIMs in equilibrium down to temperatures below the Kondo temperature $T_K,$ with the numerical error of the RAMPS scaling as a weak power law of inverse temperature, $\sim\beta^2.$ While established methods like CT-QMC allow to accurately compute the Green's function in the SIAM with unmatched efficiency, an appealing aspect of RAMPS is its applicability to impurity interactions that give rise to a severe sign problem in QMC approaches.

\section{Method}
\subsection{Model}
\label{sec:model}
In this Article, all presented results have been computed for the symmetric SIAM at half filling which is described by the Hamiltonian
\begin{align}\nonumber
\hat{H} &=  \sum_{\sigma=\uparrow,\downarrow}\sum_k \Big[ \big( t_k \hat{d}_\sigma^\dagger \hat{c}_{k,\sigma} + \text{h.c.}\big) + \epsilon_{k} \hat{c}_{k, \sigma}^\dagger \hat{c}_{k,\sigma}\Big] + \hat{H}_\text{imp},\\
\hat{H}_\text{imp} &=  \big(\epsilon_{d} + \tfrac{U}{2}\big) \sum_{\sigma=\uparrow,\downarrow} \hat{d}_\sigma^\dagger  \hat{d}_\sigma 
+ U  \big(\hat{n}_{\uparrow} -\tfrac{1}{2}\big)\big(\hat{n}_{\downarrow} -\tfrac{1}{2}\big).
\label{eq:SIAM_Hamiltonian}
\end{align}
Here, $t_k$ are hopping amplitudes between the impurity and the $k$-th bath mode, $\epsilon_k$ are quasiparticle energies of the bath fermions, $\hat{c}_{k,\sigma}^\dagger\, (\hat{c}_{k,\sigma})$ are creation (annihilation) operators for fermions at bath-mode $k$ with spin $\sigma.$ Moreover, $\hat{d}_\sigma^\dagger\, (\hat{d}_\sigma)$ creates (annihilates) a spin-$\sigma$ fermion on the impurity and $\hat{n}_\sigma = \hat{d}_\sigma^\dagger\hat{d}_\sigma$ measures the corresponding occupation number. Lastly, $\epsilon_d$ and $U$ are the impurity onsite potential and the local impurity Hubbard repulsion, respectively.
The model is at half filling for $\epsilon_d=-\tfrac{U}{2}$. The partition sum is given by $Z=\text{Tr}[\exp(-\beta \hat{H})],$ where $\beta = 1/T$ is the inverse temperature in natural units.

\subsection{Impurity observables as overlap of ``temporal wavefunctions''}
\label{sec: overlap}

\begin{figure}[h]
\begin{overpic}[width=.95\linewidth]{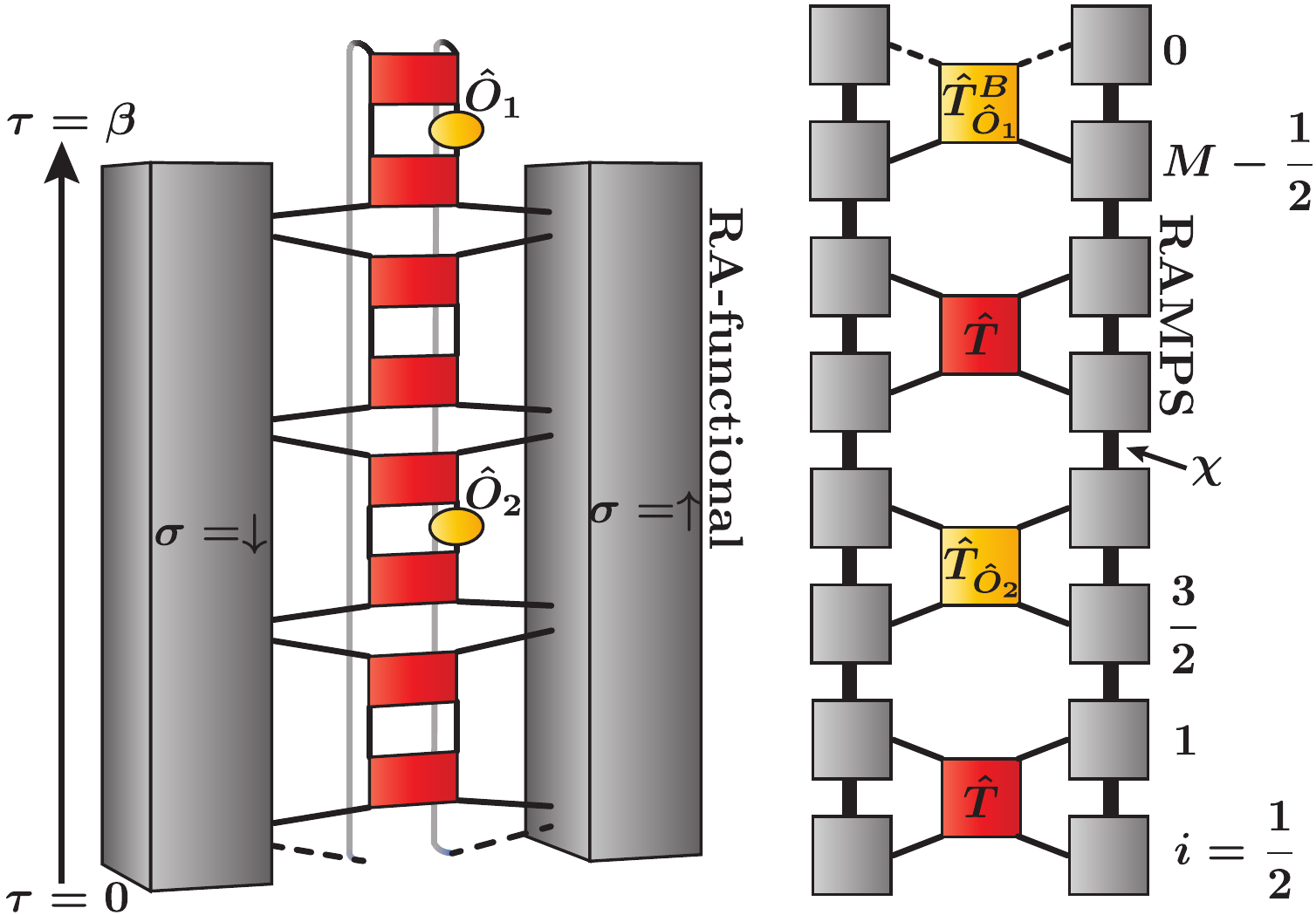}
\put(0,66){\footnotesize a)}
\put(56,66){\footnotesize b)}
\end{overpic}
\caption{Schematic representation of $\langle \hat{O}_2(\tau_n)\hat{O}_1(0)\rangle_\beta$ as a) effective path integral, Eq.\,(\ref{eq:path_integral_overlap}): Grey boxes represent the RA-functional for $\sigma = \uparrow$ and $\sigma = \downarrow,$ respectively. The local impurity evolution operator $\exp\big(-\frac{\delta\tau}{2} \hat{H}_\text{imp}\big)$ is represented by red rectangles (``gates'') and the observables $\hat{O}_1,\hat{O}_2$ are shown as yellow ovals. Antiperiodic boundary conditions (ABC) are imposed in the contraction of the last impurity gate at $\tau=\beta$
with the ingoing RA-functional variables at $\tau=0$, as indicated by dashed lines.; b) MPS-MPO contraction: The RAMPS with bond dimension $\chi$ (grey) is constructed via the Fishman-White algorithm, the local impurity gates (red, possibly including given observables) are obtained via analytical manipulations and represent two half timesteps (i.e. a full timestep) here. The physical indices of the RAMPS are labelled in half-steps with indices $i\in \{0,1/2,\dots,M-1/2 \},$ where $i=m$ and $i=(2m+1)/2$ refer to the ingoing and outgoing leg at step $m,$ respectively. The ingoing RAMPS-leg $i= 0$ (dashed line) is brought to the last position to form a MPS-MPO contraction. ABC are implemented in the last impurity gate, $\hat{T}_{\hat{O}_1}^{B}.$}
\label{fig:imag_time}
\end{figure}

We wish to evaluate  imaginary-time correlation functions of impurity observables, given by
\begin{equation}
    \langle \hat{O}_2(\tau)\hat{O}_1(0)\rangle_\beta = \frac{1}{Z}\text{Tr}\Big(e^{-(\beta-\tau) \hat{H}}\,\hat{O}_2 \,e^{-\tau \hat{H}}\,\hat{O}_1 \Big).
    \label{eq:expectationvaue_trace}
    \end{equation} 
Our method rests on the separation of the impurity and bath contributions in the path integral representation of Eq.\,(\ref{eq:expectationvaue_trace}). This allows us to represent the right-hand side as an overlap of ``temporal wavefunctions,'' which can be evaluated by constructing and contracting an efficient MPS representation of these wavefunctions. \\

    The path integral is obtained by first defining a discrete-time grid with points $\tau_m = m\,\cdot\,\delta\tau$ and $m\in[0,M].$ The parameters $\beta$ and $M$ determine the time step $\delta\tau = \beta/M.$ Then, the discrete-imaginary-time evolution operator $\hat{U}$ is defined as a second-order Trotter decomposition of $e^{-\delta\tau \hat{H}}$ into a local ``impurity part'' and a ``hybridization part'':  \begin{equation}
        \hat{U} \equiv e^{-\delta\tau/2 \hat{H}_\text{imp}} \cdot e^{-\delta\tau (\hat{H}-\hat{H}_\text{imp})} \cdot e^{-\delta\tau/2 \hat{H}_\text{imp}}
.
        \label{eq:trotter_decomposition}
         \end{equation}
         For $\tau =\tau_n,$ we can thus approximate Eq.\,(\ref{eq:expectationvaue_trace}) as
    \begin{equation}
        \langle \hat{O}_2(\tau)\hat{O}_1(0)\rangle_\beta = \langle \hat{O}_2(\tau_n)\hat{O}_1(0)\rangle_{\beta}^{\rm{discr}} +\mathcal{O}\big[(\delta\tau)^2 \big],
         \label{eq:trotter_expectation_value}
    \end{equation} where 
    \begin{equation}
        \langle \hat{O}_2(\tau_n)\hat{O}_1(0)\rangle^{\rm{discr}}_\beta \equiv  \frac{1}{\tilde{Z}} \text{Tr}\Big( \hat{U}^{M-n}\,\hat{O}_2\, \hat{U}^n\,\hat{O}_1\Big),
        \label{eq:trotter_expectation_value2}
        \end{equation} and $\tilde{Z} \equiv \text{Tr}\big(\hat{U}^M\big).$
        Dropping the Trotter error of order $(\delta\tau)^2$ in Eq.\,(\ref{eq:trotter_expectation_value}) is the only analytical approximation in the method. \\
        
To transform Eq.\,(\ref{eq:trotter_expectation_value2}) into a more tractable form, we eliminate the bath degrees of freedom using its exact path integral representation in terms of Grassmann variables. By performing the Gaussian integral over all the bath variables and making appropriate variable substitutions\,\cite{thoenniss2022efficient}, the expectation value in Eq.\,(\ref{eq:trotter_expectation_value2}) can be rewritten as
\begin{equation}
\frac{1}{\tilde{Z}}\int d(\bar{\bm{\eta}},\bm{\eta}) \, \mathcal{I}[\{ \bm{\eta}_\downarrow \}] e^{-\bar{\bm{\eta}}_\downarrow \bm{\eta}_\downarrow} \mathcal{D}^{\hat{O}_1,\hat{O}_2}_n[\bar{\bm{\eta}}_\downarrow, \bm{\eta}_\uparrow] e^{-\bar{\bm{\eta}}_\uparrow \bm{\eta}_\uparrow}\mathcal{I}[\{ \bar{\bm{\eta}}_\uparrow \}],
\label{eq:path_integral_overlap}
\end{equation} where $\bar{\bm{\eta}}_\sigma,\bm{\eta}_\sigma$ are vectors of impurity Grassmann variables on the discrete-imaginary-time grid (see App.\,\ref{app:overlap} for details).
\\

The kernel $\mathcal{D}^{\hat{O}_1,\hat{O}_2}_n$ encodes the local impurity dynamics defined by $\hat{H}_\text{imp},$ as well as the observables $\hat{O}_1,\hat{O}_2.$ The kernel $\mathcal{I}$ is the fermionic Gaussian RA-functional defined in Eq.\,(\ref{eq:IM_def}). It encodes the hybridization of impurity and bath and is fully defined by the continuous hybridization function $\Delta(\tau)$ which, in turn, is defined as the Fourier transform of Eq.\,(\ref{eq:hybridization_def}). In Eq.\,(\ref{eq:path_integral_overlap}), we exploited that the two spin species $\sigma =\uparrow,\downarrow$ do not mix in the environment: We split the RA-functional into two identical expressions, one for each species respectively, which makes the final MPS representation more efficient.  \\

Although the RA-functional is represented on a discrete-time grid, it accounts for the full continuum of temporal correlations which has been fully integrated out except for the grid points $\tau_m.$ For trivial impurity evolution $\hat{H}_\text{imp}=0$, the evolution operator $\hat{U}$ from Eq.\,(\ref{eq:trotter_decomposition}) contains no Trotter error and consequently Eq.\,(\ref{eq:path_integral_overlap}) coincides with the exact continuous-time result in Eq.\,(\ref{eq:expectationvaue_trace}), at all grid points $\tau_n$ for any choice of $\delta\tau.$
\\

We note that Eq.\,(\ref{eq:path_integral_overlap}) can formally be viewed as overlap of wavefunctions,
\begin{equation}
\langle \hat{O}_2(\tau_n)\hat{O}_1(0)\rangle_\beta = \frac{1}{\tilde{Z}}\bra{\mathcal{I}} \hat{D}_n^{\hat{O}_1,\hat{O}_2}\ket{\mathcal{I}},
\label{eq:overlap}
\end{equation} where we introduced the temporal operator- and wavefunction representation of the Grassmann kernels from Eq.\,(\ref{eq:path_integral_overlap}).
Importantly, $\hat{D}_n^{\hat{O}_1,\hat{O}_2}$ is a product operator since the local impurity evolution operator $\hat{U}_\text{imp}\equiv \exp(-\delta\tau \hat{H}_\text{imp})$ couples only neighboring points on the time-grid. All time-nonlocal effects in the impurity dynamics are induced by the bath and are therefore fully included in the RA-wavefunction $|\mathcal{I}\rangle.$ Crucially, as the latter is Gaussian, time-non-locality can efficiently be handled at an analytical level and then translated into the many-body RA-wavefunction $|\mathcal{I}\rangle.$ with established techniques. \\

Eq.\,(\ref{eq:overlap}) is pictorially represented in Fig.\,\ref{fig:imag_time} (left). The grey boxes represent the RA-functionals for $\sigma = \uparrow$ and $\sigma = \downarrow,$ respectively. The local impurity evolution operator $\hat{U}_\text{imp}$ is represented by red rectangles and the observables $\hat{O}_1,\hat{O}_2$ are shown as yellow ovals. By representing a trace as in Eq.\,(\ref{eq:expectationvaue_trace}) using Grassmann variables, antiperiodic boundary conditions (ABC) are generically introduced. The ABC are imposed in the contraction of the last impurity gate at $\tau=\beta$
with the ingoing RA-functional variables at $\tau=0$, as indicated by dashed lines.  

\subsection{Evaluating the overlap as tensor contraction} 
Since the many-body Hilbert space, in which the wavefunction $|\mathcal{I}\rangle$ is defined, is exponentially large with the number of time-grid points $M$, the overlap in Eq.\,(\ref{eq:overlap}) cannot generally be evaluated exactly in practice.
However, if the entanglement of the temporal wavefunction $|\mathcal{I}\rangle$ is moderate, one can seek an efficient representation of $|\mathcal{I}\rangle$ as matrix-product state (MPS). In Sec.\,\ref{sec:results}, we investigate the bond dimensions $\chi$ needed for an accurate MPS representation of $|\mathcal{I}\rangle$ in different physical regimes and present a study of the resulting numerical error.\\

The MPS representation can be obtained as follows: Since the RA-wavefunction represents a Gaussian Grassmann kernel, it is formally of Bardeen-Cooper-Schrieffer form, 
\begin{equation}
    |\mathcal{I}\rangle \sim \prod_{i,j=0,1/2,\dots}^{M-1/2} \Big(1+ \mathcal{G}_{ij}\hat{c}^\dagger_i \hat{c}^\dagger_j\Big)|\emptyset\rangle,
    \label{eq:Bogoliubov}
    \end{equation}
    where indices $i,j\in \{0,1/2,\dots,M-1/2 \}$ run in half-steps, such that $i=m$ and $i=(2m+1)/2$ refer to the ingoing and outgoing variable at time step $m,$ respectively. From Eq.\,(\ref{eq:Bogoliubov}) it is clear that $|\mathcal{I}\rangle$ can be obtained by applying a succession of parity-conserving rotations on the vacuum state. We exploit this property and represent these rotations as quantum gates in a circuit which, applied to the many-body vacuum and contracted, yields the MPS representation of $|\mathcal{I}\rangle.$\\

In practice, it is favorable to use only nearest-neighbor Givens- and Bogoliubov-rotations in the circuit construction. We determine such a set of rotations through the Fishman-White algorithm\,\cite{Fishman15Compression,thoenniss2022nonequilibrium} which exploits the decay of temporal bath correlations as seen by the impurity. This allows to represent $|\mathcal{I}\rangle$ by a fairly shallow circuit consisting of $\mathcal{O}(M\cdot l)$ gates, where $l$ is the localization length of the RA-wavefunction's ``natural orbitals'' in imaginary time. The action of this circuit on the vacuum can be computed at a computational cost of $\mathcal{O}\left(\chi^3 Ml\right)$, and the resulting MPS can be stored with $\mathcal{O} \left(\chi^2M  \right)$ memory. Note that this requires a lossy compression of the RA-wavefunction into an MPS if $\chi<2^{l}$, which usually is performed via a Singular Value Decomposition (SVD), keeping at most the largest $\chi$ singular vectors. Furthermore, we generally require the single-particle ``natural orbitals'' to be localized within a support $l\ll L$ only up to some precision $\epsilon_\mathrm{fw}$. These numerical parameters determine the error incurred in the conversion of the discretized RA into an MPS. When it is necessary to distinguish the latter from the Trotter error, Eq.\,(\ref{eq:trotter_expectation_value}), we will refer to the it as RAMPS error in the following.\\

Moreover, the product operator $\hat{D}_n^{\hat{O}_1,\hat{O}_2}$ is naturally represented as a matrix product operator (MPO) with bond dimension $\chi = 1.$ Hence, once the MPS representation of $|\mathcal{I}\rangle$ has been obtained, Eq.\,(\ref{eq:overlap}) can be evaluated as a tensor contraction between the two RAMPS and the impurity-MPO at a computational cost $\mathcal{O}\left(\chi^3 M\right)$. This contraction is diagrammatically represented in the right panel of Fig.\,\ref{fig:imag_time}, where the local impurity gates are analytically derived for given observables; antiperiodic boundary conditions are implemented in the last gate. In practice, it is numerically favorable to perform the MPS-MPO contraction separately for each summand of the trace from Eq.\,(\ref{eq:expectationvaue_trace}). In Fig.\,\ref{fig:imag_time} (right), this corresponds to a separate evaluation for each term in the contraction of the dashed legs. Note also that $G\left(\tau_n\right)$ can be evaluated at all $\tau_n$ with a total cost of $\mathcal{O}\left(\chi^3 M\right)$ instead of the naive $\mathcal{O}\left(\chi^3 M^2\right)$ through the use of cached partial overlaps.

\section{Results}
\label{sec:results}

 To assess the performance of the method, we compute the spin-degenerate Green's function,
 \begin{equation}
 G(\tau) = \langle \hat{d}_\sigma(\tau)\hat{d}_\sigma^\dagger(0)\rangle^\text{discr}_\beta,
 \label{eq:GF_def}
 \end{equation}
 cf. Eq.~(\ref{eq:trotter_expectation_value2}).
 [To simplify the notation in Eq.~(\ref{eq:GF_def}) and in the following, we remove the discrete time index $n$ from the variable $\tau_n$. Instead, we will represent imaginary-time arguments using the continuous variable $\tau$.]
 All results presented in this Article have been obtained for a metallic bath with a flat density of states with half-bandwidth $D,$ characterised by the hybridization function
\begin{equation}
\Delta(i\omega_n)=\frac{\Gamma}{2} \int_{-D}^{D} d\epsilon\, \frac{1}{i\omega_n - \epsilon}.
\end{equation}
In particular, we choose $D=100\,\Gamma$, which puts us close to the wide-band limit, and specify all energy and time scales in units of $\Gamma$. Within the Fishman-White algorithm for converting the single-particle bath correlation matrix to a many-body MPS, the localization length $l$ of the bath natural orbitals in imaginary time is chosen such that the mode's population is $\epsilon_\mathrm{fw}$-close to $0$ or $1$, with $\epsilon_\mathrm{fw}=10^{-12}$, with a hard upper limit of $l=14$. For singular-value decompositions performed after each application of a two-site rotation, we keep the largest $\chi$ singular values unless they are exactly 0. We verified numerically that the result is not strongly dependent on the exact choice of these parameters. Since we consider a particle-number conserving baths and impurity, we exploit this symmetry at the level of the MPS tensors. The method was implemented using the \textsc{ITensor} library\,\cite{itensor,itensor-r0.3}, building on an implementation of the Fishman-White algorithm from Ref. \cite{ItensorGaussianMPS}.\\
\begin{figure}
    \centering
    \includegraphics{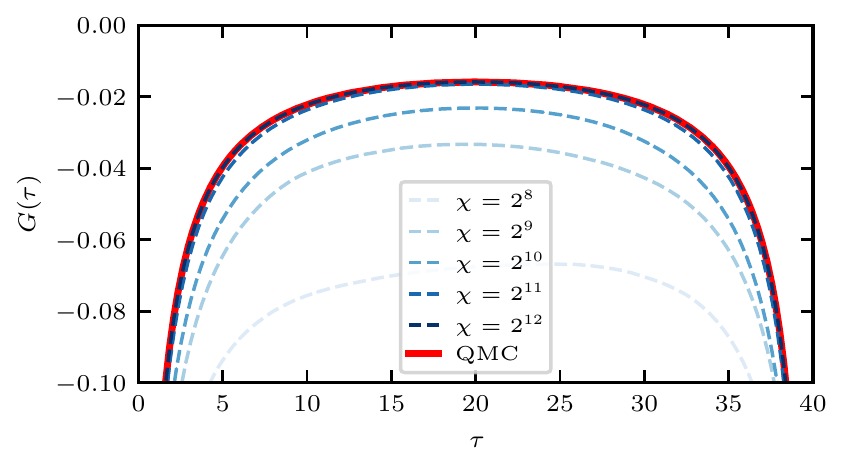}
    \caption{$G(\tau)$ for $U=4\Gamma$ and $\Gamma\beta=40$ at half filling. Dashed blue lines indicate results with a time step of $\delta\tau=1/(16\,\Gamma)$ for various bond dimension (the larger the darker the shade). The solid red line shows exact CT-QMC result, with an errorbar smaller than the linewidth.
    }
    \label{fig:Gbeta}
\end{figure}

In Fig.\,\ref{fig:Gbeta}, we report $G(\tau)$ for $\tau \in [0,\beta[$ for $\Gamma\beta=40$ and $U=4\Gamma$ using a time step of $\Gamma\delta \tau=1/16$. On the scale of Fig.\,\ref{fig:Gbeta}, the systematic Trotter error due to the finite time step is not visible in comparison to the numerically exact, discretization-error free CT-QMC result. For the smallest bond dimension considered, $\chi = 2^7 = 128$, the violation of particle-hole symmetry is evident as a result of aggressive truncation in the circuit application. However, as the bond dimension $\chi$ is increased, particle-hole symmetry is restored well before convergence of the curve $G(\tau)$ with bond dimension is achieved. Convergence in $\chi$ is reached for $\chi\approx 2^{12} = 4096$ for quantitative agreement on the scale of Fig.\,\ref{fig:Gbeta}.\\

To simplify the analysis, we will concentrate on $G(\beta/2)$ in the following. This is both numerically and physically motivated. First, the propagator can generally be expected to be least accurate around $\beta/2$ and, judging from Fig.\,\ref{fig:Gbeta}, the error at $G(\beta/2)$ is indeed a good proxy to the error over the full range of $\tau\in[0,\beta]$. Second, $-\Gamma\beta G(\beta/2)$ approaches the spectral function at zero frequency, $ \mathcal{A}(\omega = 0)$, in the limit of low temperature and is thus a physically meaningful quantity. Since $G(\beta/2)$ vanishes as $1/\beta$, we will consider the absolute deviation as an error measure in the following unless otherwise stated.\\

\begin{figure}
    \centering
    \includegraphics{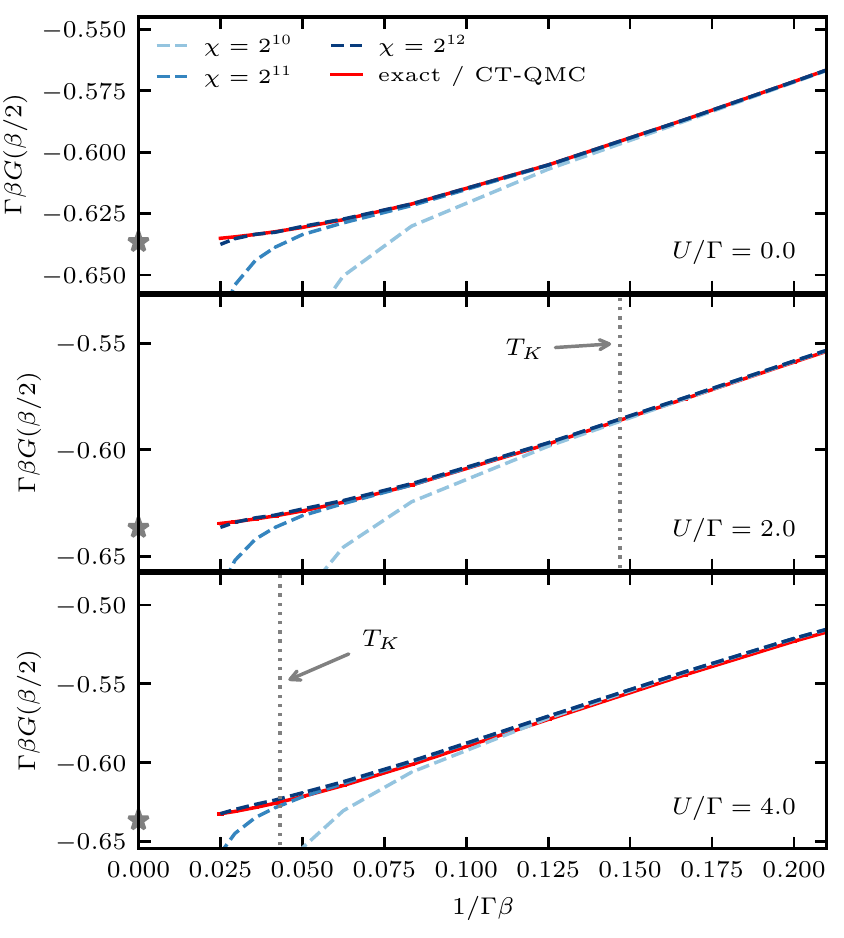}
    \caption{$\Gamma\beta G(\beta/2)$ for different values of $U=0,2\Gamma,4\Gamma$ (upper, middle and lower panel, respectively) as a function of the dimensionless temperature $1/\Gamma\beta$ at half filling. For finite $U$, the wide-band limit Kondo temperature\,\cite{BA_Tsvelick83a,BA_Tsvelick83b} $T_K=\sqrt{\frac{\Gamma U}{4}}\exp(-\frac{\pi U}{4 \Gamma})$ is indicated by the grey dotted vertical line. The zero-temperature value from the Friedel sum rule\,\cite{Langreth66Friedel}, $-\lim_{\beta\to\infty}\Gamma\beta G(\beta/2)=A(\omega = 0)=2/\pi,$ is indicated by a grey star on the left $y$-axis. Different bond dimensions are shown in different shades of blue, from $\chi=2^{10}=1024$ (light blue) to $\chi=2^{12}=4096$ (dark blue), using a time step of $\delta\tau=1/(16\,\Gamma)$. The exact result (for the noninteracting case) or numerically exact result from CT-QMC is shown as a solid red line (QMC error bars are smaller than the linewidth).}
    \label{fig:Gbetahalf}
\end{figure}

In Fig.\,\ref{fig:Gbetahalf}, we report $\Gamma\beta G(\beta/2)$ at a time step of $\Gamma \delta\tau=1/16$ for various bond dimension and interaction strengths $U/\Gamma \in\{0,2,4\}$ together with the exact result, obtained either analytically for $U=0$ or numerically from CT-QMC\,\cite{Seth26Triqs/CTHYB} for finite $U$. For $U=0,$ where Trotter errors are absent, we find that the largest bond dimensions that we considered are required to obtain reasonably converged results for the lower end of the temperatures studied. These observations carry over to finite interaction strength $U,$ up to the introduction of a systematic Trotter error, which is most visible for $U=4\,\Gamma$ in the lower panel of Fig.\,\ref{fig:Gbetahalf}. The convergence of the result with respect to bond dimension is qualitatively similar for all interaction strengths. In the following, we will thus consider the convergence with respect to bond dimension in more detail for the noninteracting case, where the RAMPS error is the only source of error, before briefly discussing the Trotter error. Note that the Kondo temperature $T_K$ obtained from a Bethe-Ansatz solution in the wide-band limit\,\cite{BA_Tsvelick83a,BA_Tsvelick83b} is indicated in Fig.\,\ref{fig:Gbetahalf}, illustrating that we are able to obtain quantitatively accurate results slightly below $T_K$ for the values of $U$ considered here.\\

\begin{figure}
    \centering
    \includegraphics{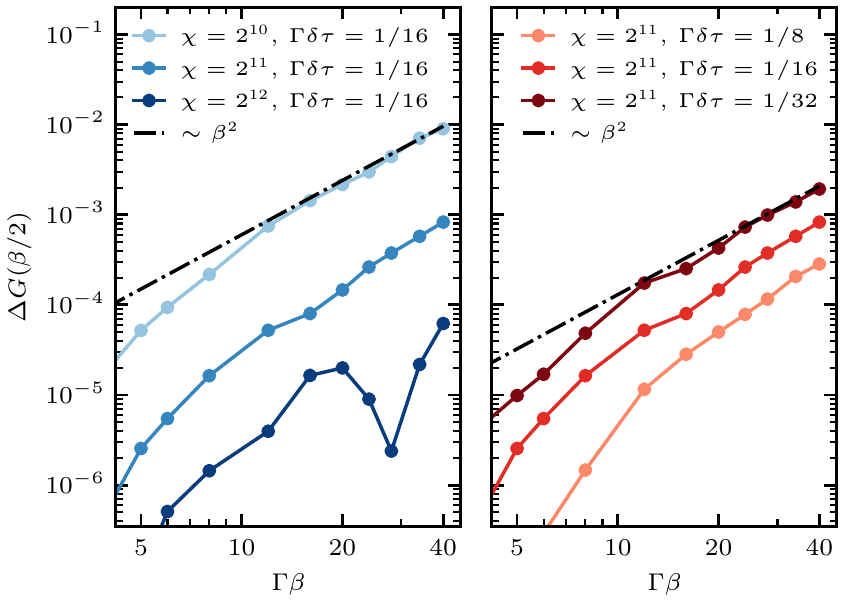}
    \caption{Absolute deviation of $ G(\beta/2)$ with respect to the exact noninteracting solution ($U=0$) as a function of $\beta$ on log-log scale. Left panel: Different shades of blue indicate bond dimensions, from $\chi=2^{10}=1024$ (lightest) to $\chi=2^{12}=4096$ (darkest) for a time step of $\delta\tau=1/(16\,\Gamma)$. Right panel: Different shades of red indicate different time steps, from largest, $\delta\tau=1/(8\,\Gamma)$ (lightest) to smallest $\delta\tau=1/(32\,\Gamma)$ (darkest) for a bond dimension of  $\chi=2048$. The black dash-dotted line is intended to serve as a guide to the eye approximating the power-law-like growth of the error with $\beta$ for larger $\beta$.}
    \label{fig:Error_Gbetahalf}
\end{figure}

The scaling of the RAMPS error as a function of $\beta$ is shown in Fig.\,\ref{fig:Error_Gbetahalf}. For a fixed bond dimension and time step, the error appears to grow as $\beta^2$ in the limit of large $\beta$, although the scaling is steeper at lower $\beta$. Increasing the bond dimension for a fixed time step suppresses the error while decreasing the time step for a fixed bond dimension increases it.\\

\begin{figure}
    \centering
    \includegraphics{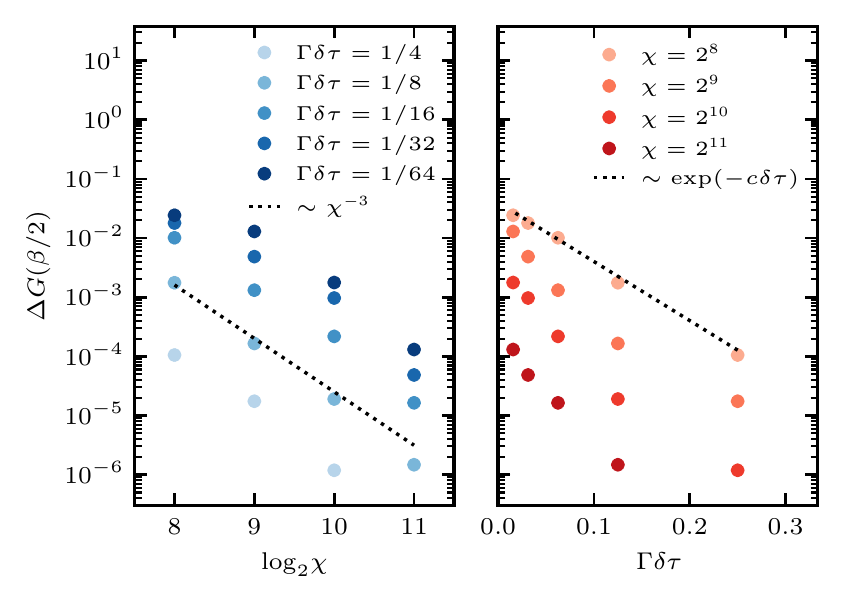}
    \caption{Behaviour of the absolute deviation of $ G(\beta/2)$ for $U=0$ from the exact result with respect to numerical parameters. The left panel shows the error as a function of $\chi$ for fixed $\beta\,\Gamma=8$ and several $\delta\tau$. In the right panel, the error is reported as a function of $\delta\tau$ for fixed $\beta\,\Gamma=8$ and several $\chi$. The dashed black lines are intended as guides to the eye and are obtained by approximating the functional dependence of the error.}
    \label{fig:Error_cuts}
\end{figure}

To understand these two behaviours better, we consider the RAMPS error as a function of the bond dimension and the time step, for a single temperature $\Gamma\beta=8$ in Fig.\,\ref{fig:Error_cuts}. For a fixed time step, the error is best described as a power law $\chi^{-\alpha}$ with $\alpha\approx 3$. Note that this is a rough estimate of the functional form, and should not be taken as a quantitative claim given the variation present in the data. Taken together with the scaling of the  error with $\beta$ of $\beta^2,$ $\chi$ should scale as $\beta^\frac{2}{3}$ for a fixed error at a fixed time step. This implies that the computational resources for a fixed error scale as $\beta^3$, where a factor of $\beta^2$ comes from the $\chi^3$ scaling of the individual tensor network contractions and an additional power of $\beta$ appears due to the linear discretization in imaginary time. On the other hand, the error for a fixed bond dimension decreases with growing time step. A reasonable fit to the error as function of timestep is obtained with $c_1\exp\left( -c_2\delta\tau\right)$ with positive constants $c_1$ and $c_2$, which implies a linearly decreasing error in the timestep for small timesteps and a finite error in the continuum limit for a fixed bond dimension.\\
\begin{figure}
    \centering
    \includegraphics{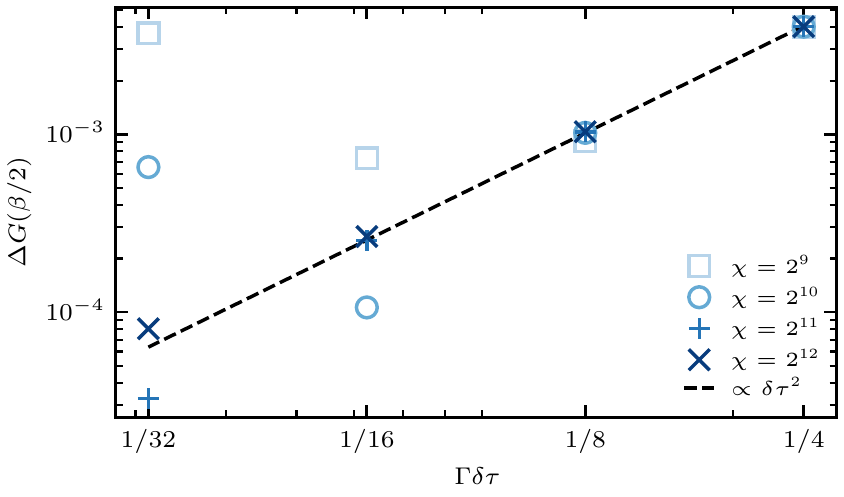}
    \caption{Absolute deviation of $G(\beta/2)$ for $U=4\Gamma$ and $\Gamma \beta=8$ from the exact result as a function of $\Gamma \delta \tau$ for various bond dimensions. The dashed black line is intended as a guide to the eye of the expected discretization error in a second-order Trotter splitting, $\mathcal{O}\big[(\delta \tau)^2\big]$.}
    \label{fig:trotter_error}
\end{figure}

The Trotter error is expected to scale as $(\delta\tau)^2$ as we employ the second-order Trotter scheme introduced in Eq.\,(\ref{eq:trotter_decomposition}). Indeed, we find that the deviation to the exact CT-QMC data is $\mathcal{O}\big[(\delta\tau)^2\big]$ in Fig.\,\ref{fig:trotter_error} as long as we consider results which are converged in the bond dimension. In this particular problem, the sign of the Trotter error and the finite bond dimension error are opposite, such that the two may cancel partially and give the impression of a lower error for smaller bond dimensions. Note that we present data for an alternative time discretization scheme with ``simultaneous'' evolution of bath and impurity in App.\,\ref{app:simultaneous_evolution}. The latter may allow for a reduced Trotter error by making a discrete-time approximation that avoids a Trotter decomposition but is instead based on taking the continuous-time limit of the full effective action only in its noninteracting part.

\section{Discussion and Conclusion}
In this Article, we present a detailed numerical analysis of the RAMPS approach in imaginary time for a metallic bath. The approach is based on a matrix product state representation of the retarded action which is naturally formulated in terms of the bath hybridization function.\\
We used the RAMPS approach to compute the imaginary-time Green's function $G(\tau)$ of an interacting impurity in the wide-band limit, and demonstrated that we can retain computational accuracy down to the Kondo regime for all parameter sets investigated.\\

Moreover, we explored the dependence of the error on finite bond dimension $\chi$ and Trotter step $\delta\tau.$ Our results are consistent with a polynomial complexity, which we tentatively estimate as $\mathcal{O}(\beta^3)$, of the method in imaginary time for a metallic bath, i.e. critical bath correlations. This is in line with expectations from prior studies on the real-time axis\,\cite{thoenniss2022efficient,thoenniss2022nonequilibrium}, as well as the intuition that critical correlations lead to violation of the area-law of entanglement entropy, both in space and in (euclidean) time. \\

We observed signs of convergence in the time step $\delta\tau$ for fixed bond dimension $\chi,$ suggesting that this limit is well-defined and attainable in principle. By separately examining the numerical errors of RAMPS and Trotter approximation, we establish that there exists a temperature range where both errors are controlled.  Given reasonable numerical resources, this temperature range extends into the Kondo regime for the moderate interaction strengths considered. We find that reducing the time step amplifies the RAMPS error. At the same time, decreasing $\delta\tau$  mitigates the Trotter error. Consequently, selecting an appropriate $\delta\tau$ becomes crucial for achieving a balance between these errors and accurately computing observables at low temperatures.\\

Based on these results, we conclude that the RAMPS approach is well suited to compute impurity properties, including higher-order correlators, to a good accuracy with numerical resources that scale polynomially in the inverse temperature and target accuracy. An extension to multi-orbital QIMs is formally straightforward. One way to achieve this is to partition the baths associated with different orbitals between left and right RAMPS, with an impurity MPO that is adjusted accordingly. In the case of diagonal impurity-bath couplings, this orbital separation is exact, while for non-diagonal couplings, it introduces an additional Trotter error of $\mathcal{O}\big[(\delta\tau)^2\big]$. Whether multi-orbital QIMs are numerically tractable then depends crucially on whether the bond dimension $\tilde{\chi}$ of the multi-orbital RAMPS saturates the theoretical upper bound of $\chi^n$ for an $n$-orbital impurity. While we do not expect the bound to be saturated in physically realistic scenarios, this question should be addressed in future work.\\

In light of the excellent performance of this method for real-time QIMs following a quench\,\cite{thoenniss2022nonequilibrium,thoenniss2022efficient}, an exciting avenue emerges by combining the tools from both the real-time and imaginary-time domains. This integration could offer an opportunity to efficiently compute non-equilibrium quantum impurity problems on the full L-shaped Keldysh contour. This remains a challenging task, as there are limited methods available that can achieve both efficiency and accuracy \cite{PhysRevB.96.155126}. \\

In particular, one promising application is to the formulation of non-equilibrium DMFT which involves the hybridization function on the L-shaped contour. As the RAMPS and its influence-functional counterpart in real-time are directly defined by the hybridization function, the combined approach is naturally suited to tackle such non-equilibrium scenarios. Extending our method to this domain may overcome existing limitations and provide a powerful tool for studying the dynamics and transport properties of QIMs under non-equilibrium conditions.\\

\begin{acknowledgements}
 We thank G. Chan, G. Mazza, N. Ng, G. Park, D. Reichman, and L. Tagliacozzo for discussions. Support by the European Research Council (ERC)
under the European Union’s Horizon 2020 research and innovation program
(grant agreement No.~864597) and by the Swiss National Science Foundation is
gratefully acknowledged. The Flatiron Institute is a division of the Simons Foundation.
J.T. thanks A. Georges and the CCQ for their hospitality and resources provided during the preparation of this Article.
\end{acknowledgements}
\bibliography{main}

\appendix
\clearpage
\newpage

\section{Overlap of temporal wavefunctions}
\label{app:overlap}
The exact path integral representation of Eq.\,(\ref{eq:trotter_expectation_value2}) is obtained in the standard way: We decompose all exponentials into infinitesimal time steps and insert identity resolutions of Grassmann coherent states, $\mathds{1}=\int d(\bar{\eta},\eta)\,e^{-\bar{\eta} \eta}|\eta\rangle\langle \bar{\eta}|$, between all operator multiplications. Since the environment evolution is Gaussian, all bath variables can be integrated out. After making appropriate variable substitutions, this yields Eq.\,(\ref{eq:path_integral_overlap}) and (\ref{eq:overlap}), here restated for convenience:
\begin{multline*}
    \langle \hat{O}_2(\tau_n)\,\hat{O}_1(0)\rangle\\=
    \frac{1}{\tilde{Z}}\hspace{-0.1cm}\int \hspace{-0.15cm}d(\bar{\bm{\eta}},\bm{\eta}) \mathcal{I}[\{ \bm{\eta}_\downarrow \}] e^{-\bar{\bm{\eta}}_\downarrow \bm{\eta}_\downarrow} \mathcal{D}^{\hat{O}_1,\hat{O}_2}_n[\bar{\bm{\eta}}_\downarrow, \bm{\eta}_\uparrow] e^{-\bar{\bm{\eta}}_\uparrow \bm{\eta}_\uparrow}\mathcal{I}[\{ \bar{\bm{\eta}}_\uparrow \}] \\
= \bra{\mathcal{I}_\downarrow} \hat{D}_n^{\hat{O}_1,\hat{O}_2}\ket{\mathcal{I}_\uparrow}.
\end{multline*}
Here, 
\begin{align*}
\bm{\eta}_\sigma\equiv &(\eta_{\sigma,0}, \eta_{\sigma,1/2},\hdots,\eta_{\sigma,M-1/2}),\\ \bar{\bm{\eta}}_\sigma\equiv & (\bar{\eta}_{\sigma,M}, \bar{\eta}_{\sigma,1/2},\hdots,\bar{\eta}_{\sigma,M-1/2})\\
d(\bar{\bm{\eta}},\bm{\eta}) \equiv & \prod_\sigma d\bar{\eta}_{\sigma,M}d\eta_{\sigma,0}\,d\bar{\eta}_{\sigma,1/2} \, d\eta_{\sigma,1/2}\\
&\times \prod_{m=1}^{M-1} d\bar{\eta}_{\sigma,m} \, d\eta_{\sigma,m}d\bar{\eta}_{\sigma,m+1/2} \, d\eta_{\sigma,m+1/2}.
\end{align*}
The kernel encoding the local impurity evolution reads:
\begin{widetext}
\begin{multline}
\mathcal{D}^{\hat{O}_1,\hat{O}_2}_n[\bar{\bm{\eta}}_\downarrow, \bm{\eta}_\uparrow] =\prod_{\substack{m=1\\ m\neq n}}^{M-1}\exp\Big[- \Big(\eta_{\uparrow,m}\eta_{\uparrow,m-1/2}  + \bar{\eta}_{\downarrow,m}\bar{\eta}_{\downarrow,m-1/2}  + \delta\tau \mathcal{H}_\text{imp}[\{-\eta_{\uparrow,m},\eta_{\uparrow,m-1/2},\bar{\eta}_{\downarrow,m},-\bar{\eta}_{\downarrow,m-1/2} \}]\Big)\Big]\\
\times \tilde{\mathcal{O}}_2[-\eta_{\uparrow,n},\eta_{\uparrow,n-1/2},\bar{\eta}_{\downarrow,n},-\bar{\eta}_{\downarrow,n-1/2}]\, \tilde{\mathcal{O}}_1[\eta_{\uparrow,M},\eta_{\uparrow,M-1/2},-\bar{\eta}_{\downarrow,M},-\bar{\eta}_{\downarrow,M-1/2}],\label{eq:kernel_imp_full}
\end{multline}
\end{widetext}
where $\tilde{\mathcal{O}}_{1,2}$ are the combined kernels of the operator $\hat{O}_{1,2}$ and the evolution gate at the corresponding time, respectively.

Moreover, we have defined the imaginary-time RA-functional:
\begin{equation}
\mathcal{I}[\{\bm{\eta}\}] \equiv \exp\Big[ \sum_{m,n=0}^{M-1}\eta_{m+1/2}\Big(-(\delta\tau)^2\Delta_{m,n} +\delta_{m,n}\Big) \eta_{n}\Big].
\label{eq:IM_def}
\end{equation}

As stated in Eq.\,\eqref{eq:overlap}, the manipulated path integral can be directly read off as a sandwich between states and operators in the time domain. 
Indeed, note that the ``barred'' and ``non-barred'' variables have been suitably renamed to make this sandwich expression manifest.
Our goal is to compactly represent such states and operator as MPSs and a MPO, respectively, and hence to compute the sandwich via a standard tensor contraction. 
Notice that the normalization of $\ket{I}$ cancels out when dividing by $\tilde{Z}$, so we can tacitly assume the state to be normalized by $\sqrt{\tilde{Z}}$ and drop the denominator.
\\
\section{Operator representation of the path integral in Eq.\,(\ref{eq:path_integral_overlap})}
\label{app:MPS-conversion}
The RA-vector is uniquely determined by Eq.\,(\ref{eq:IM_def}), and can be abstractly written in the form $\mathcal{I}[\{\bm{\eta}\}] = \exp\big( \bm{\eta}^T\mathcal{G} \bm{\eta}\big)$ where $\mathcal{G}$ is an antisymmetric matrix.  
The mapping from Grassman function to many-fermion wavefunction works by straightforward replacement of Grassmann variables with corresponding creation operators on the vacuum.
Thus, a Gaussian Grassmann function $\mathcal{I}[\{\bm{\eta}\}] = \exp\big( \bm{\eta}^T\mathcal{G} \bm{\eta}\big)$ can be straightforwardly associated with a Gaussian, BCS-type many-body wavefunction 
\begin{equation}
\ket{I}= \exp\big((\hat{\bm{c}}^\dag)^T \mathcal{G}\hat{\bm{c}}^\dag \big)\ket{\emptyset}, \quad\hat{\bm{c}}^\dag = (\hat{c}^\dagger_{0},\hat{c}^\dagger_{1/2},\hat{c}^\dagger_{1},\hdots,\hat{c}^\dagger_{M-1/2}).\label{eq:many-body_wf}
\end{equation}  

Such a many-body wavefunction is entirely determined by its correlation matrix, which is the input to the Fishman-White algorithm that we use to determine the MPS representation of $\ket{I_\uparrow}$. Note that because of our choice of conventions above, $\bra{I_\downarrow}$ is represented by the transposed vector (no complex conjugation), and hence is literally the same MPS. \\

The final ingredient is the operator $\hat{D}_n^{\hat{O}_1,\hat{O}_2}$ which corresponds to the Grassmann kernel $\mathcal{D}_n^{\hat{O}_1,\hat{O}_2}$ in Eq.\,(\ref{eq:kernel_imp_full}) and acts on the many-body Fock space in the temporal domain.
Since the impurity action in Eq.\,(\ref{eq:kernel_imp_full}) is local in time, the operator $\hat{D}_n^{\hat{O}_1,\hat{O}_2}$ is a product operator:
\begin{equation}
\hat{D}_n^{\hat{O}_1,\hat{O}_2} = \underbrace{\hat{T} \otimes \hdots \otimes\hat{T}}_{(n -1)\text{ times}} \otimes \hat{T}_{\hat{O}_2} \otimes \underbrace{\hat{T} \otimes \hdots \otimes\hat{T}}_{(M-n-1) \text{ times}} \otimes \hat{T}_{\hat{O}_1}^B.
\end{equation}
With reference to Fig.\,\ref{fig:imag_time}, here each $\hat{T}$ is the ``temporal-domain-version'' of the impurity evolution operator $\exp(-\delta\tau \hat{H}_\text{imp})$, acting between a ``$\uparrow$'' two-fermion space (originally corresponding to the tensor product of input and output Hilbert spaces of the ``$\uparrow$'' impurity fermion) and a ``$\downarrow$'' two-fermion space (originally corresponding to the tensor product of input and output Hilbert spaces of the ``$\downarrow$'' impurity fermion).
$\hat{T}_{\hat{O}_2}$ is the ``temporal-domain-version'' of the impurity evolution operator including the observable operator $\hat{O}_2$. Finally, $\hat{T}^B_{\hat{O}_1}$ is the same for the observable $\hat{O}_1$ up to a slight modification to take into account the antiperiodic boundary conditions.\\

\section{Evaluating the RA-functional}
\label{app:computing_IF}

In this section, we sketch how to construct the RA-functional, Eq.\,(\ref{eq:IM_def}), for a given hybridization function $\Delta(i\omega_n),$ which we obtain either explicitly from a given spectral density or as the output of a previous DMFT cycle.
In the former case, it is given by:
\begin{equation}
    \Delta(i\omega_n) = \int \frac{d\epsilon}{2\pi} \,\Gamma(\epsilon)\frac{1}{i\omega_n - \epsilon + \mu},
    \label{eq:hybridization_def}
\end{equation} where $\Gamma(\epsilon)=2\pi\sum_k t_kt_k^* \,\delta(\epsilon-\epsilon_k)$ is the spectral density, $t_k$ are hopping amplitudes between the impurity and the $k$-th bath mode, $\epsilon_k$ are bath energies and $\mu$ is the chemical potential of the bath. The noninteracting impurity Green's function is defined by the Matsubara sum:
\begin{equation}
    G^0(\tau) = \frac{1}{\beta}\sum_{n} e^{-i\omega_n\tau}\, G^0(i\omega_n),
      \label{eq:non_interacting_impurity_GF_fourier}
\end{equation} where 
$$G^0(i\omega_n) = \frac{1}{i\omega_n - \epsilon_d -\Delta(i\omega_n)}.$$

Note that Eq.\,(\ref{eq:non_interacting_impurity_GF_fourier}) is given in the convention where the impurity onsite potential $\epsilon_d$ is included in the RA-functional --- if one chooses to define it as part of the impurity, $\epsilon_d$ has to be set to zero here. In this case, for $\epsilon_d\neq 0,$ Eq.\,(\ref{eq:overlap}) would contain a Trotter error even for $U=0.$\\

The Grassmann kernel of the RA-functional has the form: 
\begin{equation}
    \label{eq:IF_form}
    \mathcal{I}[\{\bm{\eta}\}] \equiv \exp\Big[\sum_{m,n=0}^{M-1}\eta_{m+1/2}\, \mathcal{G}_{m,n}\, \eta_n\Big].
    \end{equation}
    The components $\{\mathcal{G}_{m,0}\}$ can be written as multipoint correlation functions in the noninteracting continuous-time problem:
\begin{widetext}
\begin{align}
    \mathcal{G}_{m,0}& = \int d\big(\bar{\eta}_\tau,\eta_\tau \big)
    \exp\Big[-\int_0^\beta d\tau \,\bar{\eta}_\tau \partial_\tau \eta_\tau -\int_0^\beta d\tau\int_0^\beta d\tau^\prime \,\bar{\eta}_\tau\Delta(\tau-\tau^\prime)\,\eta_{\tau^\prime}\Big] \eta_{\tau_M} \bar{\eta}_{\tau_{M-(m+1)}}\prod_{\substack{l=1\\ l\neq M-(m+1)}}^{M-1} \eta_{\tau_l}\bar{\eta}_{\tau_l}.
\end{align}
\end{widetext}
Here, $\Delta(\tau)$ is the conventional hybridization function defined by its spectral representation.
Using Wick's theorem, one can rewrite $\mathcal{G}_{m,0}$ as the determinant of a matrix containing the noninteracting Green's function at the time-points $\tau_n.$
For this, we define the matrix:
\begin{equation}
    \bm{G}^0 \equiv \begin{pmatrix}
    G^0(0) &  G^0(\delta\tau) &   \dots & G^0(\beta)\\
     G^0(-\delta\tau) &  G^0(0) &    \dots & G^0(\beta-\delta\tau)\\
      G^0(-2\delta\tau) &  G^0(-\delta\tau) &   \dots & G^0(\beta-2\delta\tau)\\
      \vdots & \vdots  & \ddots\\
      G^0(-\beta) & G^0(-\beta + \delta\tau) & \dots & G^0(0)
    \end{pmatrix}.
\end{equation}
The components $\mathcal{G}_{m,0}$ for $m<M$ are then given by:
\begin{align}
\frac{\mathcal{G}_{m,0}}{\mathcal{Z}}=
    \det \Big[{\bm{G}^0}_{\big|_{[0,1,\dots,\cancel{m+1},\dots,M-1], [m+1,1,\dots,\cancel{m+1},\dots,M-1]}}\Big].
\end{align}
For $m=M,$ we have:
\begin{align}
\frac{\mathcal{G}_{M,0}}{\mathcal{Z}}=
    -  \det \Big[{\bm{G}^0}_{\big|_{[0,1,\dots,M-1], [M,1,\dots,M-1]}}\Big].
\end{align}
Here, we introduced a (sign-adjusted) partition sum, $$\mathcal{Z} = -1/ \det\Big[{\bm{G}^0}\big|_{[0,\dots,M-1],[0,\dots,M-1]} \Big],$$ where we defined the minus sign to cancel the minus sign included in the definition of $G^0.$
Since $G^0(\tau) = - G^0(\tau+\beta),$ we can always evaluate the Green's function with a time argument in the range $\tau\in [0,\beta].$ Furthermore, note that because of $\mathcal{G}_{m,n} = -\mathcal{G}_{n+M,m},$ Eq.\,(\ref{eq:IF_form}) is fully determined by the $M$ different values $\{\mathcal{G}_{m,0}\}.$

\section{Simultaneous evolution of impurity and environment}
\label{app:simultaneous_evolution}
Here, we explain an alternative second-order discrete time approximation which avoids successive evolution of bath and impurity, thereby avoiding systematic shifts in observables as a result of the time-discretization error. We refer to it as the ``simultaneous evolution'' scheme in contrast to the ``successive evolution'' scheme introduced in the main text, Eq.\,(\ref{eq:trotter_decomposition}). Again, we start by defining a discrete-time grid with time step $\delta \tau,$ analogously to Sec.\,\ref{sec: overlap}.
Rather than making a Trotter decomposition as in Eq.\,(\ref{eq:trotter_decomposition}), we define the evolution operator as $\hat{U}=\exp\big( -\delta \tau \,\hat{H}\big),$ where $\hat{H}$ is the full Hamiltonian of bath and impurity, Eq.\,(\ref{eq:SIAM_Hamiltonian}). 

\begin{figure}
\begin{overpic}[width=.95\linewidth]{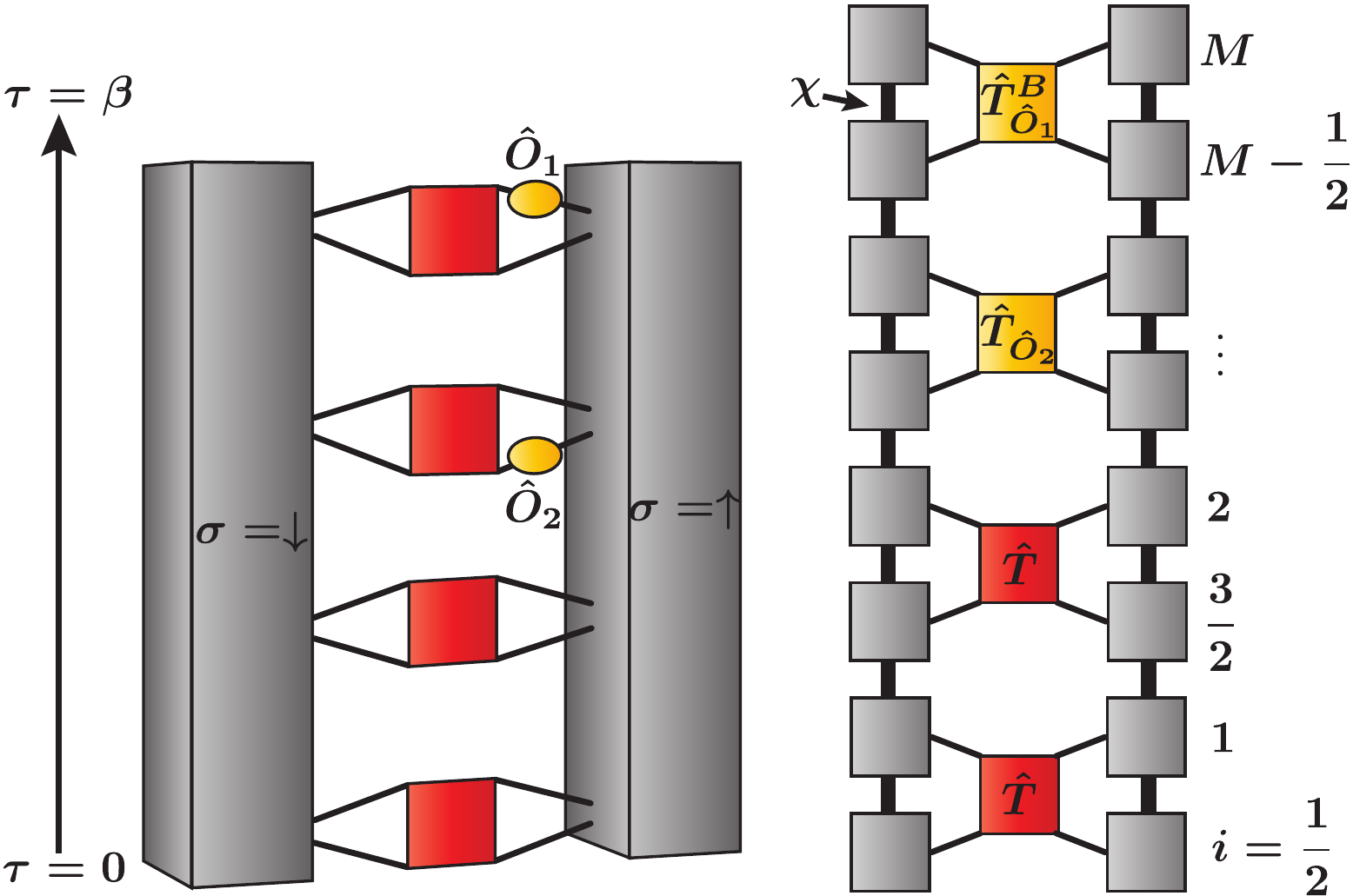}
\put(0,63){\footnotesize a)}
\put(57,63){\footnotesize b)}
\end{overpic}
\caption{Schematic representation of $\langle \hat{O}_2(\tau_n)\hat{O}_1(0)\rangle_\beta$ for ``simultaneous'' evolution as a) path integral, Eq.\,(\ref{eq:overlap_cont_time}): Grey boxes represent the RA-functionals for $\sigma = \uparrow$ and $\sigma = \downarrow,$ respectively. The local impurity evolution operator $e^{-\delta\tau \hat{H}_\text{imp}}$ is represented by red rectangles and the observables $\hat{O}_1,\hat{O}_2$ are shown as yellow ovals. As opposed to the Trotter scheme from Eq.\,(\ref{eq:trotter_decomposition}) used for results in the main text, the output variables of the impurity at $\tau =\beta$ are connected to the output variables of the RA-functional at $\tau = \beta$ with antiperiodic boundary conditions; and as b) MPS-MPO contraction: The RAMPS with bond dimension $\chi$ (grey) is obtained via the Fishman-White algorithm, the local impurity gates are analytically derived for given observables and boundary conditions are included in the last gate. The physical indices of the RAMPS are labelled in half-steps with indices $i\in \{1/2,\dots,M-1/2,M \},$ where $i=(2m+1)/2$ and $i=m+1$ refer to the ingoing and outgoing leg at step $m,$ respectively. No reordering of legs is necessary here. Antiperiodic boundary conditions are absorbed in the last impurity gate, $\hat{T}_{\hat{O}_1}^{B}.$}
\label{fig:imag_time_simult}
\end{figure}

The approximation can formally be written in the form of Eq.\,(\ref{eq:trotter_expectation_value}), with the thermal expectation value given by:
\begin{align}\nonumber
&\langle \hat{O}_2(\tau_n)\hat{O}_1(0)\rangle_\beta\\
&=\frac{1}{Z}\int d(\bm{\eta}_\uparrow,\bm{\eta}_\downarrow) \, e^{S_\text{hyb}[\bm{\eta}_\downarrow]}\mathcal{D}^{\hat{O}_1,\hat{O}_2}_n[\bm{\eta}_\downarrow, \bm{\eta}_\uparrow] e^{S_\text{hyb}[\bm{\eta}_\uparrow]},
\label{eq:expec_simul}
\end{align}
with
\begin{widetext}
\begin{align}
\nonumber
\mathcal{D}^{\hat{O}_1,\hat{O}_2}_n[\bm{\eta}_\downarrow, \bm{\eta}_\uparrow] =& \prod_{\substack{m=1\\ m\neq n+1}}^{M-1}\exp\Big[- \Big(\eta_{\uparrow,m}\eta_{\uparrow,m-1/2}  + \eta_{\downarrow,m}\eta_{\downarrow,m-1/2}  + \delta\tau \mathcal{H}_\text{imp}[\{-\eta_{\uparrow,m},\eta_{\uparrow,m-1/2},\eta_{\downarrow,m},-\eta_{\downarrow,m-1/2} \}]\Big)\Big]\\
&\times \tilde{\mathcal{O}}_2[-\eta_{\uparrow,n+1},\eta_{\uparrow,n+1/2},\eta_{\downarrow,n+1},-\eta_{\downarrow,n+1/2}]\, \tilde{\mathcal{O}}_1[\eta_{\uparrow,M},\eta_{\uparrow,M-1/2},-\eta_{\downarrow,M},-\eta_{\downarrow,M-1/2}]
\label{eq:impuritykernel_full_conttime}\\
S_\text{hyb}[\bm{\eta}_\sigma] =& (\delta\tau)^2\,\Bigg[ \sum_{m = 1}^{M-1}  \sum_{n = 1}^{M} \eta_{\sigma,m} \Delta_{m,n}\eta_{\sigma,n-1/2} - \sum_{n = 1}^{M} \eta_{\sigma,M } \Delta_{M,n}\eta_{\sigma,n-1/2} + \sum_{m=1}^{M - 1} \eta_{\sigma,m} \eta_{\sigma,m+1/2}\Bigg] + \eta_{\sigma,M}\eta_{\sigma,1/2},
\label{eq:s_hyb_contime}
\end{align}
\end{widetext}
and 
\begin{align}\nonumber
\bm{\eta}_\sigma\equiv &( \eta_{\sigma,1/2},\hdots,\eta_{\sigma,M}),\\\nonumber
d(\bm{\eta}_\uparrow,\bm{\eta}_\downarrow) \equiv& \prod_\sigma d\eta_{\sigma,M}\,d\eta_{\sigma,1/2}\,\prod_{m=1}^{M-1} d\eta_{\sigma,m} \, d\eta_{\sigma,m+1/2},\\\nonumber
Z&=\text{Tr}[\exp(-\beta \hat{H})].
\end{align}
Here, $\tilde{\mathcal{O}}_{1,2}$ are the combined kernels of the operator $\hat{O}_{1,2}$ and the evolution gate at the corresponding time, respectively. We made manipulations to the path integral in such a way that the sign convention in the impurity kernel $\mathcal{D}$ in Eq.\,(\ref{eq:impuritykernel_full_conttime}) is unchanged with respect to Eq.\,(\ref{eq:kernel_imp_full}). Eq.\,(\ref{eq:expec_simul}) is schematically shown in Fig.\,\ref{fig:imag_time_simult} (left). The elements of the hybridization function $\Delta_{m,n}$ in Eq.\,(\ref{eq:s_hyb_contime}) are related to the matrix elements $\mathcal{G}_{m,n}$ from Eq.\,(\ref{eq:IF_form}) by $
\mathcal{G}_{m,n} = -(\delta \tau)^2 \, \Delta_{m,n} + \delta_{m,n}.$\\

Note that Eq.\,(\ref{eq:expec_simul}) does not have the form of an overlap. To obtain an equation that can interpreted as overlap, we introduce a more convenient notation and rewrite $S_\text{hyb}$ as \begin{equation}
    S_\text{hyb}[\bm{\eta}_\sigma] =\frac{1}{2}\sum_{m,n=1}^M\begin{pmatrix}
        \eta_{\sigma,m-1/2}\\
         \eta_{\sigma,m}
    \end{pmatrix}^T \mathbf{A}_{m,n}\begin{pmatrix}
        \eta_{\sigma,n-1/2}\\
         \eta_{\sigma,n}
    \end{pmatrix},
\end{equation} where the matrix $\mathbf{A}$ has the following subblocks:
  \begin{equation}\nonumber
    \mathbf{A}_{m,n} = 
    \begin{cases}
     +(\delta \tau)^2 \, \bm{\Delta}_{m,n} + \begin{pmatrix}
    0 & -\delta_{m,n+1} \\
    \delta_{m+1,n} & 0
    \end{pmatrix}& (a)  \\
      -(\delta \tau)^2 \, \bm{\Delta}_{M,n} + \begin{pmatrix}
    0 & -\delta_{M-1,n} \\
    0 & 0
    \end{pmatrix}  & (b) \\
     -(\delta \tau)^2 \, \bm{\Delta}_{m,M} + \begin{pmatrix}
    0 & 0 \\
    \delta_{M-1,m} & 0
    \end{pmatrix} & (c) \\
      -(\delta \tau)^2 \, \bm{\Delta}_{M,M}  & (d)\\
     -(\delta \tau)^2 \, \bm{\Delta}_{1,M} + \begin{pmatrix}
    0 & -1 \\
    0 & 0
    \end{pmatrix} & (e) \\
     -(\delta \tau)^2 \, \bm{\Delta}_{M,1} +\begin{pmatrix}
    0 & 0 \\
    1 & 0
    \end{pmatrix}& (f).
    \end{cases}
    \end{equation}
   Here, we distinguish the cases:
   \begin{align}\nonumber
       (a) & \quad 1\leq m,n<M,\\\nonumber
       (b) & \quad m=M,\, 1<n<M,\\\nonumber
       (c) & \quad 1<m< M,\, n=M, \\\nonumber
       (d) & \quad  m=n= M, \\\nonumber
       (e) &\quad m=1,\, n= M, \\\nonumber
       (f) & \quad m=M,\, n= 1,
   \end{align}
    and we have defined
    $$\bm{\Delta}_{m,n} \equiv \begin{pmatrix}
        0 & -\Delta_{n,m}\\
        \Delta_{m,n} & 0
    \end{pmatrix}.$$
Introducing a new set of impurity variables $\bm{\zeta}_\sigma$ (and thus doubling the degrees of freedom), we can rewrite 
\begin{align}\nonumber
    e^{ S_\text{hyb}[\bm{\eta}_\sigma]} &= \frac{1}{pf(\mathbf{A}^{-1})} \int d\bm{\zeta}_\sigma \exp\Big[-\bm{\zeta}_\sigma^T \bm{\eta}_\sigma
   + \frac{1}{2} \bm{\zeta}_\sigma^T \mathbf{A}^{-1} \bm{\zeta}_\sigma\Big] \\\nonumber
   &= \frac{1}{pf(\mathbf{A}^{-1})} \int d\bm{\zeta}_\sigma \exp\Big[-\bm{\eta}_\sigma^T \bm{\zeta}_\sigma
   + \frac{1}{2} \bm{\zeta}_\sigma^T \mathbf{A}^{-1} \bm{\zeta}_\sigma\Big],
\end{align}
where the second line is obtained from the first line by substituting $$\begin{pmatrix}
    \zeta_{\sigma,m-1/2}\\
    \zeta_{\sigma,m}
\end{pmatrix} \to \begin{pmatrix}
    -\zeta_{\sigma,m-1/2}\\
    -\zeta_{\sigma,m}
\end{pmatrix},$$ which flips the sign in the first term of the exponential and introduces a trivial prefactor $(-1)^{2M}=1.$
By inserting this into Eq.\,(\ref{eq:expec_simul}), we obtain the expectation value in overlap form:
\begin{align}\nonumber
\langle \hat{O}_2(\tau_n)&\hat{O}_1(0)\rangle_\beta\\\nonumber
\propto \int &d(\bm{\eta}_\uparrow,\bm{\eta}_\downarrow)\int d(\bm{\zeta}_\uparrow,\bm{\zeta}_\downarrow)  \\
&\tilde{\mathcal{I}}[\bm{\zeta}_\downarrow]\, e^{-\bm{\eta}_\downarrow ^T \bm{\zeta}_\downarrow } \, \mathcal{D}^{\hat{O}_1,\hat{O}_2}_n[\bm{\eta}_\downarrow, \bm{\eta}_\uparrow] \,e^{-\bm{\zeta}_\uparrow^T \bm{\eta}_\uparrow}\,
   \tilde{\mathcal{I}}[\bm{\zeta}_\uparrow],
\label{eq:overlap_cont_time}
\end{align}
with \begin{equation}
    \tilde{\mathcal{I}}[\bm{\zeta}_\sigma] = \exp\Big[\frac{1}{2}\, \bm{\zeta}_\sigma^T\mathbf{A}^{-1}\bm{\zeta}_\sigma \Big].\label{eq:IF_continuous}
    \end{equation}
At this point, Eq.\,(\ref{eq:overlap_cont_time}) has the same form as Eq.\,(\ref{eq:overlap}) and we can therefore evaluate the overlap as usual, see Fig.\,\ref{fig:imag_time_simult}.\\

\begin{figure}
    \centering
    \includegraphics{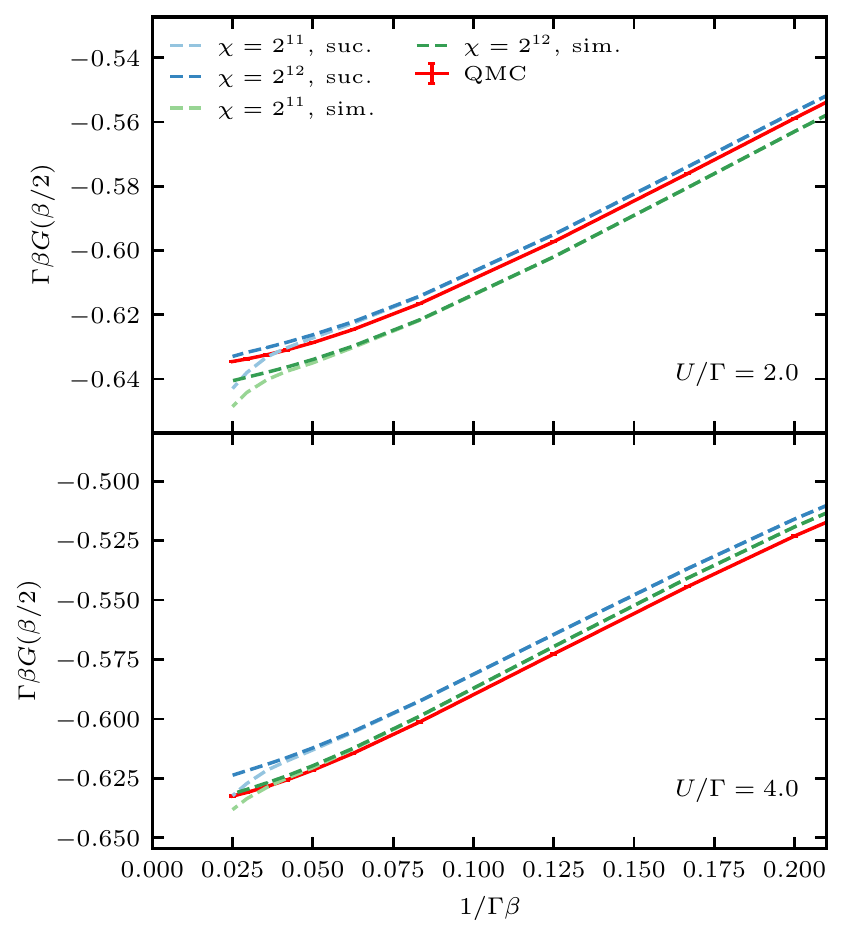}
    \caption{$\beta G(\beta/2)$ as a function of $1/\beta$, obtained using the simultaneous (sim., green dashed lines) and successive (suc., blue dashed lines) time-discretization scheme with a time step of $\delta\tau=1/(8\,\Gamma)$ for both. Different shades indicate different bond dimensions, and the upper (lower) panel shows results for $U=2$ ($U=4$). The numerically exact result from CT-QMC is shown in red.}
    \label{fig:Gbetahalf-simultaneous}
\end{figure}

\begin{figure}
    \centering
    \includegraphics{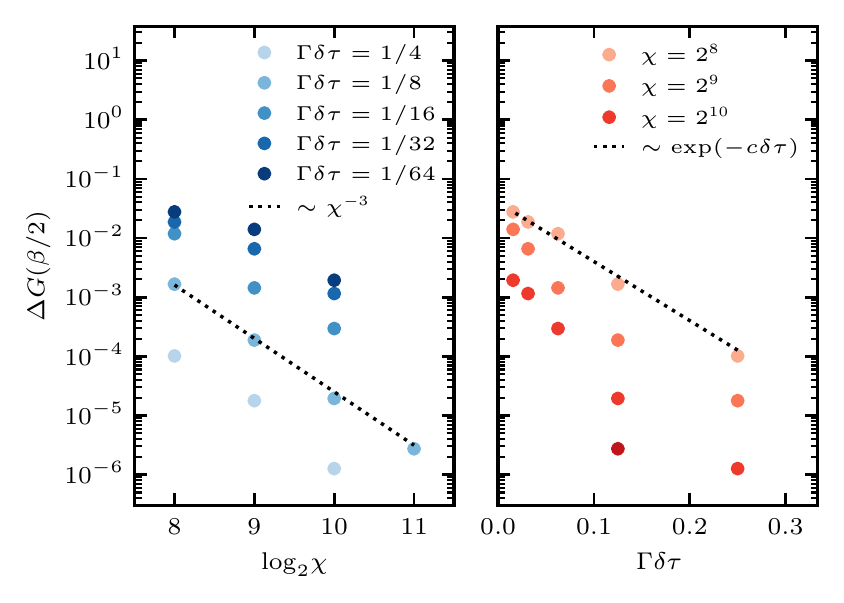}
    \caption{Behaviour of the absolute deviation of $\beta G(\beta/2)$ for $U=0$ from the exact result with respect to numerical parameters for the simultaneous evolution. The left panel shows the error as a function of $\chi$ for fixed $\beta\,\Gamma=8$ and several $\delta\tau$. In the right panel, the error is reported as a function of $\delta\tau$ for fixed $\beta\,\Gamma=8$ and several $\chi$. The dashed black lines are intended as guides to the eye extracting the approximate functional dependence of the error in terms of the numerical parameters.}
    \label{fig:errGbetahalfcuts_sim}
\end{figure}

For the chosen time step $\delta\tau = 1/(16\,\Gamma)$ as shown in Fig.\,\ref{fig:Gbetahalf}, both discretization schemes produce highly comparable outcomes. However, when examining a larger time step of $\delta\tau = 1/(8\,\Gamma)$ in Fig.\,\ref{fig:Gbetahalf-simultaneous}, a notable contrast emerges between the two schemes. At $U = 2\Gamma$, the Trotter scheme, as discussed in the main text, demonstrates superiority, whereas the ``simultaneous scheme'' exhibits significantly improved accuracy for $U = 4\Gamma$, particularly at lower temperatures.\\

Examining the RAMPS error in Fig.\,\ref{fig:errGbetahalfcuts_sim} at $U = 0$, we find similar orders of magnitude and qualitative convergence behavior to that of the second-order Trotter scheme (``successive evolution'') that is depicted in Fig.\,\ref{fig:Error_cuts}. Thus, the observed discrepancy in accuracy illustrated in Fig.\,\ref{fig:Gbetahalf-simultaneous} can be attributed to the analytically distinct discrete-time approximation schemes, rather than to the numerical error of the RAMPS.\\

While the conventional Trotter approximation benefits from a transparent $(\delta\tau)^2$-scaling of the time-discretization error, we currently lack a rigorous error theory for the ``simultaneous scheme.'' However, solely based on numerical results, no clear preference exists for either scheme at present. It would be intriguing to develop a better understanding of the error behavior in the ``simultaneous scheme'' which may enable calculations in the strong coupling regime (large $U$) using larger time steps $\delta\tau$ than the ones that are required in the second-order Trotter scheme. Such an advancement could potentially yield substantial reductions in required numerical resources. Nonetheless, our current findings provide inconclusive evidence in this regard.

\newpage

\end{document}